\documentclass[sigconf]{acmart}

\AtBeginDocument{%
  \providecommand\BibTeX{{%
    \normalfont B\kern-0.5em{\scshape i\kern-0.25em b}\kern-0.8em\TeX}}}

\copyrightyear{2024}
\acmYear{2024}
\setcopyright{acmcopyright}\acmConference[MSR 2024]{21st International Conference on Mining Software Repositories}{April 15-16, 2024}{Lisbon, Portugal}

\usepackage{booktabs}
\usepackage{tikz,lipsum}
\usepackage[most]{tcolorbox}
\usepackage{multirow}
\usepackage{setspace}

\begin{document}


\title{Investigating the Utility of ChatGPT in the Issue Tracking System: An Exploratory Study}


\author{Joy Krishan Das}
\affiliation{%
  \country{University of Saskatchewan, Canada}
    }
\email{joy.das@usask.ca}

\author{Saikat Mondal}
\affiliation{%
  \country{University of Saskatchewan, Canada}
    }  
\email{saikat.mondal@usask.ca}

\author{Chanchal K. Roy}
\affiliation{%
  \country{University of Saskatchewan, Canada}
  }
\email{chanchal.roy@usask.ca}


\begin{abstract}
Issue tracking systems serve as the primary tool for incorporating external users and customizing a software project to meet the users' requirements. However, the limited number of contributors and the challenge of identifying the best approach for each issue often impede effective resolution. Recently, an increasing number of developers are turning to AI tools like ChatGPT to enhance problem-solving efficiency. While previous studies have demonstrated the potential of ChatGPT in areas such as automatic program repair, debugging, and code generation, there is a lack of study on how developers explicitly utilize ChatGPT to resolve issues in their tracking system. Hence, this study aims to examine the interaction between ChatGPT and developers to analyze their prevalent activities and provide a resolution. In addition, we assess the code reliability by confirming if the code produced by ChatGPT was integrated into the project's codebase using the clone detection tool NiCad. Our investigation reveals that developers mainly use ChatGPT for brainstorming solutions but often opt to write their code instead of using ChatGPT-generated code, possibly due to concerns over the generation of ``hallucinated'' code, as highlighted in the literature.

\end{abstract}


\maketitle

\section{Introduction}

\begin{figure}[!t]
    \centering
    \includegraphics[width=0.45\textwidth]{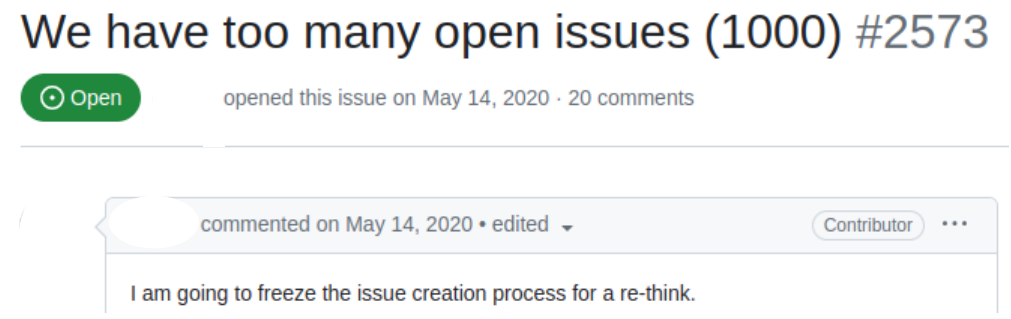}
    \caption{The project maintainer has temporarily halted the creation of new issues due to the high number of unresolved ones.}
    \label{fig:issue-overload}
\end{figure}

The involvement of the key developers with the external community is a crucial factor in ensuring the long-term sustainability of open source software (OSS) projects \cite{west2005contrasting}.  Given the open nature of these systems, users commonly express their requirements by framing them as features or issues to be incorporated into future versions of the software product \cite{heppler2016cares}. Moreover, within open-source GitHub projects, it is normal for any individual to initiate new issues in the project's issue tracker. However, as the volume of issues surpasses the project contributors' capacity (as shown in Figure~\ref{fig:issue-overload}), it is reasonable to infer that not all issues receive adequate attention. Some may be closed without resolution or implicitly overlooked \citep{kikas2015issue}. A significant challenge that programmers frequently encounter, irrespective of the issue's scale and complexity, is determining the most appropriate strategy for resolution \cite{abbas2023incorporating}. However, a clear understanding of the problem often reveals better pathways. Such pathways can resolve the reported issues within the designated time and budget. 
%
Therefore, introducing ChatGPT into such situations can help the developer provide several answers to each query, unveiling diverse ways to resolve the issue at hand. 
The promise of ChatGPT in automatic program repair \cite{surameery2023use}, bug fixing \cite{haque2023potential}, and conceptualization \cite{filippi2023measuring} is studied and acknowledged in the literature.

\begin{figure*}
    \centering
    \includegraphics[scale=0.55]{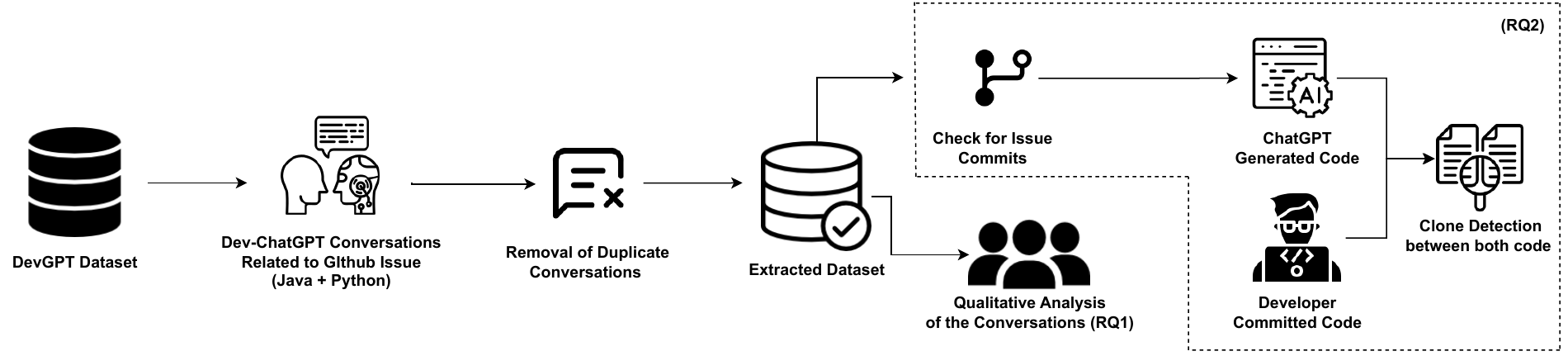}
    \captionsetup{skip=-1pt}  
    \caption{Methodology of our study.}
    \label{fig:enter-label}
\end{figure*}

While prior research has delved into the efficacy of ChatGPT across various software development tasks, there remains an unexplored realm regarding how developers leverage ChatGPT to advance in addressing issues or implementing features within an issue-tracking system. Furthermore, it is imperative to investigate whether developers trust the code snippets generated by ChatGPT and seamlessly incorporate them into the project's codebase.

In this study, we aim to explore the interactions between developers and ChatGPT found in the DevGPT dataset \cite{devgpt}. Our objective is to uncover the practical ways in which developers employ ChatGPT to enhance the efficiency of issue resolution. We also incorporate a clone detection tool called NiCad \cite{cordy2011nicad} to check if the  ChatGPT-generated code is actually incorporated into the committed files for that issue in the tracking system.

In particular, we answer two research questions (RQs) and make two contributions to this study.

\begin{itemize}
    \item[] \textbf{RQ1:How is ChatGPT used in issue-tracking systems, and what is its primary use?}
    This RQ provides insights into ChatGPT's multifaceted applications within the issue-tracking system. Furthermore, it illuminates the most frequent tasks on which developers depend on ChatGPT's aid. Therefore, by outlining ChatGPT's main application in issue-tracking systems, researchers can pinpoint possible areas for enhancement and suggest innovative methodologies to better support developers in efficiently resolving issues.

    \smallskip
    \item[] \textbf{RQ2: What is the level of developers' reliance on ChatGPT-generated code?} Given ChatGPT's reported tendency to generate hallucinated code \cite{alkaissi2023artificial, bang2023multitask}, this question attempts to understand the extent to which developers incorporate ChatGPT's code in their actual coding projects versus primarily using it for other purposes.
\end{itemize}

\smallskip
\noindent\textbf{Replication Package} can be found in our online appendix \citep{replicationPackage}
 


\section{Methodology}

Figure~\ref{fig:enter-label} presents the methodology of our study, detailing the activities in the subsequent subsections.

\subsection{Dataset Construction}

The summary of the extracted dataset is given in Table~\ref{tab:summary}. In this study, we use all the issue-sharing snapshots from DevGPT \cite{devgpt}. The DevGPT dataset is periodically expanded using the same conversation links, with each snapshot generated on a certain day. Therefore, it contains a lot of duplicate issue conversations. We thus removed these duplicate conversations with our custom script to extract data pertinent to our study. A total of $98$ distinct URLs were identified where Python was the primary language used in the code snippets generated by ChatGPT, while 27 distinct URLs were found where Java was the primary language used in code snippets. Next, we develop an additional script designed to navigate through the unique issue URL identified in the dataset to determine whether an issue is associated with a commit. Upon discovering a commit SHA within the issue URL, we proceed to extract the code corresponding to that commit into a file designated as "developer". Following this, we extract code generated by ChatGPT from the conversation link embedded in the same issue URL, storing it in a file termed ``ChatGPT''. This is done to answer our RQ2.

\subsection{Data Analysis}

To answer RQ1, we conduct a qualitative analysis of the conversations between the developer and ChatGPT and categorize them. The first two authors of this paper take part in this manual analysis and independently label the conversations. The approach unveiled five primary categories encompassing code generation, code understanding, debugging, ideation, and validation. The first and second authors exhibited a significant level of Cohen Kappa\cite{cohen1960coefficient} agreement (97.6\%) and deliberated on any discrepancies in code assignment. We persisted with this procedure until the discussion and remarks from developers on the issue ceased to uncover any other categories, reaching a state of theoretical saturation. 

To answer RQ2, we employ the popular clone detector, NiCad \cite{cordy2011nicad}, to find the near-miss clones between the code committed by the developer and the code generated by ChatGPT, specifically on the 15 issues that contain commit SHA. Subsequently, the first author further manually investigates to determine whether the ChatGPT code has been integrated into the project's codebase.

\begin{table}
    \centering
    \caption{Summary of the extracted dataset}
    \label{tab:summary}
    \begin{tabular}{lr} \toprule
         \textbf{Item} & \textbf{Value}\\ \midrule
         \# of total unique issues(Java, Python) & 125\\
         \# of issues closed & 54\\
         \# of issues open & 71\\
         \# of issues with commits & 15 \\ \bottomrule   
    \end{tabular}
    
\end{table}

\section{Study Findings}

In this section, we present our findings in accordance with RQ1 and RQ2.


\subsection{Uses of ChatGPT in the  issue tracking system (RQ1)}

Our analysis of DevGPT conversations uncovered that developers employed ChatGPT in mainly \emph{five} different ways to advance towards finding a solution to the issue. 
 
\subsubsection{ChatGPT as a code generation tool.}
Our analysis revealed that developers frequently employ ChatGPT to request code snippets, particularly when addressing feature enhancement issues. This was observed in 11 out of 14 issues tagged as feature enhancement. Since studies have explored the ability of ChatGPT to generate source code \cite{sobania2023analysis, qadir2023engineering, george2023review}, many developers are using it to code faster (shown in Table~\ref{fig:enter-label}). We have observed that around four developers express great satisfaction with the code snippets generated by ChatGPT in a single attempt. On the contrary, some developers were unable to obtain usable code even with a few prompts, as ChatGPT consistently generated similar code across all prompts. Furthermore, we have observed that developers employ ChatGPT to generate code that is not intricate but rather extensive and laborious, such as scripting a JSON converter. For example, a developer commented,

\textit{``We should just build a json converter if its too hard to rewire this. Let me get chatgpt to do it''}

\subsubsection{ChatGPT as an issue comprehension tool.}
Contributors to a project know intricate structures, nested functionalities, and specific nuances within the codebase. Whereas ChatGPT lacks awareness of the entire codebase, which may lead to occasionally `hallucinating' interpretations that diverge from the intended issue resolutions \cite{zhang2023language}. This is because when faced with ambiguous queries, ChatGPT tends to make assumptions about the developer's intentions instead of seeking clarification. This may be a strong reason why not many developers are using ChatGPT as an issue comprehension tool, as shown in Table~\ref{fig:enter-label}.
\subsubsection{ChatGPT as a debugging tool.}
We noticed developers engage in an interaction with the model, describing the symptoms of a bug, providing contextual information, and seeking guidance on potential solutions. This conversational approach fosters a dynamic debugging process that enables developers to articulate and refine their understanding of the issue. However, only a few developers used it as a debugging tool in our study dataset. This is due to the intricate nature of modern codebases, which often pose challenges in identifying the root cause of bugs. In debugging, developers usually prompt ChatGPT with the buggy code and ask the chatGPT to find out if there is any problem in the code. Such as,

\textit{``is this buggy?
<code>''}

\subsubsection{ChatGPT as an ideation tool.}
Multiple studies\cite{kojima2022large, york2023evaluating} have shown the ability of ChatGPT to serve as a collaborative partner for brainstorming solutions to challenges or sparking fresh ideas. Therefore, it can enhance creativity as it can act as a collaborative partner when brainstorming and ideation. One of the primary benefits of ChatGPT to developers is its 24/7 availability. Even in our study, a lot of developers mainly used ChatGPT for idea creation, providing guidance to resolve an issue, or improving any features. Similarly, some developers mentioned that, despite providing broken code, it gave a lot of hints on how to resolve the issue. For example,

\textit{``Built this quick prototype with the help of ChatGPT...ChatGPT transcript that got me here <Chat link>''}

\subsubsection{ChatGPT as a validation tool.}
We noticed that after a developer recommends a solution in the comments of an issue tracking system, they verify its accuracy and viability in a conversation with ChatGPT. This incorporation of ChatGPT conversations for validation marks a shift in how developers approach the verification of their suggestions. They share recorded conversations, which serve as a transparent and informative trail, allowing contributors to follow the evolution of ideas, the validation process, and the collective decision-making journey. However, few developers use ChatGPT to verify their remarks (as shown in Table~\ref{fig:enter-label}), maybe due to concerns about the AI solution's dependability \cite{zhang2023language}.

\begin{tcolorbox}[enhanced,title=\(RQ_1\) Summary,
attach boxed title to top center=
{yshift=-3mm,yshifttext=-1mm},
boxed title style={size=small}]
Developers commonly rely on ChatGPT as a beneficial collaborator during the ideation phase to address the issue at hand. This includes presenting different approaches, recommending code snippets, and detailing potential resolutions. 
\end{tcolorbox}


 \begin{table}
     \caption{Different types of activities and related developer comments after interactions with ChatGPT}
     \centering
     \begin{tabular}{llr}\toprule
          \textbf{Tasks} & \textbf{Representative Quotes} & \textbf{Freq}\\ \midrule 
          \begin{tabular}{@{}l@{}} Generation \end{tabular} & \begin{tabular}{@{}l@{}}``\textit{I used ChatGPT code interpreter to}\\\textit{ write the recursive tree functions''}\end{tabular} &  39\\\hline
          \begin{tabular}{@{}l@{}} Comprehen-\\sion\end{tabular} &\begin{tabular}{@{}l@{}}\textit{``I noticed hfind uses binary search of}\\\textit{hash databases...I discussed with}\\\textit{ChatGPT''} \end{tabular}&  9\\\hline
          \begin{tabular}{@{}l@{}} Debugging \end{tabular} &\begin{tabular}{@{}l@{}}\textit{``I'm curious if <file> is buggy which}\\\textit{is my guess and gpt-4's guess as well''} \end{tabular} &  6\\\hline
          Ideation & \begin{tabular}{@{}l@{}}``\textit{I used ChatGPT to help me get to this}\\\textit{point...It gave me enough hints that I}\\\textit{could figure out how to build a non-}\\\textit{broken version''}\end{tabular} & 61\\ \hline
          Validation & \begin{tabular}{@{}l@{}}``\textit{Here is chatgpt chat ... you may}\\\textit{increase Batch size from 100 to 20000}\\\textit{ for example depending on your GPU.''}\end{tabular}&10\\ \bottomrule
     \end{tabular} 
    \label{tab:my_label}
 \end{table} 

\subsection{Developers' dependence on ChatGPT-generated code (RQ2)}

We first determine the number of issues that were resolved after interacting with ChatGPT, assuming that these codes provided some guidance in resolving the issues. This is consistent with our previous RQ1, in which developers stated that ChatGPT helped them comprehend the procedure for developing a functional version. The statistics about the number of issues closed and open are presented in Table~\ref{tab:Statistics of Issues Closed and Open}. Significantly, the average prompt for resolved issues is smaller than that for unresolved issues, suggesting that developers might have found solutions with a minimum prompt in ChatGPT. The same pattern observed in the mean prompts for both Python and Java provides additional evidence for the correlation between prompt behavior and issue state.

In order to further explore the dependence on ChatGPT code, we analyze the 15 issues that encompass the commit SHA with NiCad. The NiCad clone detector is a tool created by Cordy and Roy \cite{cordy2011nicad} that identifies code similarities and duplication in software.
Our NiCad analysis reveals that there are 14 out of 15 issues where the developer's code and ChatGPT's code do not match. However, in one case, NiCad detects a type-1 clone with a 100\% similarity, indicating a direct replication of ChatGPT's code into the codebase. Additionally, a manual review confirmed ChatGPT's code had no match with the committed code for the remaining 14 issues. These results imply developers avoid using ChatGPT's code due to its lack of verification and potential for generating erroneous code.

\begin{tcolorbox}[enhanced,title=\(RQ_2\) Summary,
attach boxed title to top center=
{yshift=-3mm,yshifttext=-1mm},
boxed title style={size=small}]
Developers engaging with ChatGPT show quicker resolutions in closed issues with fewer prompts. Although they consult ChatGPT for ideas, developers often prefer their own code, reinforced by NiCad's finding of discrepancies in 14 out of 15 comparisons between developers and ChatGPT code.
\end{tcolorbox}
\begin{table}
    \centering
        \caption{Statistics of issues closed and open}
        \resizebox{3.4in}{!}{%
            \begin{tabular}{ccccc}\toprule
             \multirow{2}{1.5cm}{\textbf{Language}}& \multicolumn{2}{c}{\textbf{Closed}}&\multicolumn{2}{c}{\textbf{Open}} \\\cmidrule(lr){2-3} \cmidrule(lr){4-5} 
             &  \textbf{Count}&  \textbf{Mean Prompt}&  \textbf{Count}& \textbf{Mean Prompt}\\\midrule
             Python &  42&  6.83&  56& 7.34\\
             Java &  12&  3.33&  15& 4.4\\ \bottomrule
        \end{tabular}
        }
     \label{tab:Statistics of Issues Closed and Open}

\end{table}
\section{Related Works}
Many researchers have studied ChatGPT, leveraging the GPT-3.5 architecture because of its advanced language understanding and generation capabilities, signaling the potential to revolutionize fields like software engineering.  Thus, Liu et al.  \cite{liu2023refining} assess ChatGPT's code generation for 2,033 tasks, finding varied accuracy across 4,066 Java and Python instances. The study notes both high correctness and instances of errors, alongside maintainability challenges. However, it points to ChatGPT's potential for self-improvement and further AI code generation advancements.
Similarly, Tian et al.  \cite{tian2023chatgpt} investigated ChatGPT as an automated programming assistant, evaluating its code generation, repair, and summarization against benchmarks. While effective for common tasks, its limitations in sustained attention highlight the critical role of prompt engineering.


Despite its productivity potential and speculative advantages over human engineers, there's scant empirical support for these claims, particularly concerning efficiency, fairness, and safety metrics. So, Nathalia et al. \cite{10.5555/3615924.3615927} conducted an empirical study comparing ChatGPT with programmers of varying expertise on coding challenges, which reveals ChatGPT's proficiency with basic and intermediate tasks but not outperforming experts, highlighting the importance of integrating AI with human skills for enhanced collaboration and task allocation.
Conversely, the escalating intricacy of software necessitates more efficient debugging tools. Haque et al. \cite{haque2023potential} delve into ChatGPT's viability as such a tool, scrutinizing its advantages and limitations while offering insights on its seamless integration into the software development lifecycle.

\section{Discussion}

\textbf{ChatGPT aids in issue-resolving ideation.} ChatGPT's integration with issue-tracking systems has become vital in software maintenance. The iterative process of ChatGPT-driven talks enhances the clarity and strength of notions. Although ChatGPT does not possess autonomous problem-solving capabilities, its contextual understanding enables it to function as a valuable tool for developers, assisting in generating a wide range of ideas and proposing viable approaches to problem-solving.  Its guidance is crucial in directing developers, ultimately saving significant time and effort that is usually dedicated to extensive internet searches or reviewing documentation. Thus, the growing adoption of ChatGPT by developers signifies a marked shift in addressing and resolving issues, moving away from the traditional, time-intensive practices of extensive internet searches and documentation review \cite{abdullah2022chatgpt}.

\textbf{ChatGPT code is rarely utilized as it is.} 
Our research findings indicate that developers refrain from incorporating the code created by ChatGPT directly into the project's codebase. One key reason is that code generation tools like ChatGPT excel at handling straightforward code logic. However, grasping large and complicated Software Engineering projects can present numerous obstacles \cite{castelvecchi2022chatgpt}. Furthermore, the project module exhibits significant interdependence with other modules, which ChatGPT may disregard entirely when producing code \cite{castelvecchi2022chatgpt}. On the other hand, the existence of ethical issues with ChatGPT's code, which may include gender and racial bias \cite{chatterjee2023new}, could also be a significant factor contributing to the lack of reliability in the created code.

\section{Threats to Validity}

Our open-coding approach, however thorough, may be susceptible to individual biases. Although we evaluated all distinct issues in the DevGPT dataset, the small number of issues investigated may limit the wider applicability of our results.
We used NiCad to identify code similarities between developer code and ChatGPT-generated code. While NiCad is very efficient, we remain aware of the possible risks involved. So, we supplement our methodology with manual validation.

\section{Conclusion}

This paper explores the various ways employed by developers while utilizing ChatGPT within the context of an issue-tracking system. The findings indicate that ChatGPT is mostly utilized as a collaborative ally for generating ideas and finding answers. Due to its hallucinatory nature \cite{zhang2023language}, it is less commonly employed for validating developers' suggestions or remarks. Furthermore, a minimal number of developers utilize it for the purpose of issue comprehension and debugging due to ChatGPT's tendency to overlook the interconnection between several modules within a large project. 
Our findings also demonstrate that although ChatGPT is utilized for code generation in cases mostly involving feature enhancements, developers do not largely rely on the code it produces. Instead, they assimilate concepts and approaches to choose the optimal course of action and subsequently write their own code. Our study offers valuable insights into the utilization of ChatGPT as a tool for issue resolution. In our future work, we aim to extend our findings by analyzing a larger dataset that encompasses interactions between developers and ChatGPT, enabling a more extensive extrapolation of our research within issue-tracking systems.

\section*{Acknowledgement}
This research is supported in part by the Natural Sciences and Engineering Research Council of Canada (NSERC) Discovery Grants program, and by the industry-stream NSERC CREATE in Software Analytics Research (SOAR). 

\bibliographystyle{IEEEtran}
\bibliography{references.bib}

\appendix

\end{document}